\documentclass[amssymb,amsfonts,aps,prb,twocolumn,superscriptaddress]{revtex4}

\usepackage{graphicx}
\usepackage{hyperref}
\usepackage{amsmath}
\usepackage[dvipsnames]{xcolor}

\newcommand{\expESRSTM}{ manassen1989direct,durkan2002electronic,komeda2008,reviewbalatsky2012,
Mullegger14,Baumann_Paul_science_2015}
\newcommand{\theoryESRSTM}{balatsky2002esr,ortiz03,Caso14}

\usepackage{dcolumn}
\usepackage{bm}
\begin{document}

\title{Exchange mechanism for electron paramagnetic resonance of individual adatoms}

\author{  J. L. Lado}
\affiliation{
 QuantaLab, International Iberian Nanotechnology Laboratory (INL),
Av. Mestre Jos\'e Veiga, 4715-330 Braga, Portugal
}
\author{A. Ferr\'on}
\affiliation{
Instituto de Modelado e Innovaci\'on Tecnol\'ogica (CONICET-UNNE) and Facultad de Ciencias Exactas,
Naturales y Agrimensura, Universidad Nacional del Nordeste, Avenida Libertad 5400, W3404AAS Corrientes, Argentina.
}
\author{J. Fern\'andez-Rossier}
\affiliation{
 QuantaLab, International Iberian Nanotechnology Laboratory (INL),
Av. Mestre Jos\'e Veiga, 4715-330 Braga, Portugal
}
\affiliation{
 Departamento de F\'isica Aplicada, Universidad de Alicante,  03690 Spain}

\date{\today}

\begin{abstract}
We propose a new universal mechanism that makes it possible to drive 
an individual atomic spin using a spin polarized 
scanning tunnel microscope (STM) with an oscillating
electric signal.
We show that the combination of the distance dependent
exchange with the magnetic tip 
and the electrically driven mechanical oscillation of the surface spins permits 
to control their quantum state. 
  Based on a combination of density functional theory and multiplet calculations, we 
  show that  the proposed mechanism  is essential to account for the recently observed
electrically driven paramagnetic spin resonance (ESR) of 
 an individual Fe atom on a MgO/Ag(100)
  surface. Our findings set the foundation to deploy  the ESR-STM quantum sensing technique  to a much broader class of systems.

\end{abstract}

 \maketitle

\section{Introduction}

Scanning tunneling microscopy (STM)  and electron paramagnetic resonance (EPR)
are two very powerful experimental techniques whose integration has
been pursued in the last 3 decades\cite{\expESRSTM} and motivated a substantial
body of  theoretical work\cite{\theoryESRSTM}. 
In EPR, spin transitions are excited with an ac field that permits to resolve
spin excitation with a resolution limited by the intrinsic spin relaxation
broadening  of the species. In continuous wave (cw) EPR this can be down to 
a few MHz 
for amorphous hydrogenated silicon
\cite{mitchell2013x}.     However, standard detection techniques based on
induction require probing at least 10$^{7}$ spins\cite{blank2003high}, and have
thereby a very poor spatial resolution.  In contrast,   STM permits to probe
individual atoms with an exquisite spatial resolution, but when it comes to
perform spin spectroscopy,   it  relies on inelastic electron
tunneling\cite{Hirjibehedin_Lin_Science_2007} (IETS),
whose spectral resolution is
limited\cite{Jaklevic_Lambe_prl_1966} by $5.4 k_BT$, where $T$ is the
temperature.  Thus,  even for the coldest STM so far\cite{assig201310},    the
spectral resolution of IETS spectroscopy  would be above  30 GHz, ie, 3 orders
of magnitude worse off than cw-EPR.    

In a recent experimental breakthrough, Baumann
{\em et al.} \cite{Baumann_Paul_science_2015} have reported the measurement of the electron paramagnetic
resonance of an individual Fe atom deposited on top of an   atomically thin MgO
layer grown on Ag(001),   
using an spin polarized STM  tip  (see figure 1) to both drive the atom with an
$ac$ signal and to probe the resulting reaction.   
For the driving, they applied 
 a radio frequency (RF)  voltage $V_{RF}$ across the tip-sample  with frequency $f$.
The resulting change in $dc$ current, $I_{DC}$ as a function of $f$ displayed a
very narrow ($3$ MHz) resonance peak, at the frequency $f_0$ that matches the
Zeeman splitting of the magnetic adatom ground state doublet
{($26$ GHz for $B_z=0.2$ T)}.  The peak,  well
above the noise level,   would shift upon application of a magnetic field, making
it possible thereby to detect 50$\mu T$ variations with subatomic resolution.   
 
 The  experiment of Baumann {\em et al.} \cite{Baumann_Paul_science_2015},
electrically driven paramagnetic
spin resonance (ESR),
outperforms the spectral resolution of IETS-STM  spectroscopy by 4 orders of
magnitude, at the same temperature,  and  reaches the absolute detection limit,
by probing a single spin.   The recently reported application of  this remarkable setup 
to probe the magnetic moment of individual atoms nearby \cite{Choi2017,Natterer2017}  
demonstrates the potential of 
ESR-STM technique as an extremely versatile  quantum sensing tool. 

In this paper we address a fundamental question that begs for an answer in
order to understand 
the working principles of any STM-ESR setup, namely, 
how an RF voltage can drive the atomic spin.
Baumann  {\em et al.} \cite{Baumann_Paul_science_2015} proposed a mechanism that combines  two ingredients.
First, the RF electric field induces  a mechanical oscillation  $z(t)$
of the surface atom. Second, the induced modulation of the crystal field,
combined with the
spin-orbit interaction of the $d$ electrons of the surface atom,
results in transitions between the two lowest energy levels 
of the atomic spin.  Whereas the first ingredient applies for any charged
surface atom,  the second is only valid for the specific symmetry of the Fe/MgO
system.

Here we propose an alternative universal mechanism that permits
  to drive the spin of a charged surface atom, using an RF electrical voltage
and an
STM tip with a magnetic atom in the apex.  The mechanism is based on the
notion that  spin  interactions  between
the tip and the surface atom depend strongly on their distance.
The electric modulation of the surface atom position results in a variation
of the spin-spin interaction that can efficiently drive the surface spin. 
{This
changes the occupation 
of the surface spin states, that changes its average magnetic moment.  Because
of the spin-polarized nature of the STM tip, 
this leads to a magnetoresitve
change of the $dc$ current.\cite{Baumann_Paul_science_2015}}
{
In the present manuscript we focus on the nature of the coupling
between the AC voltage and the surface spin, without trying to evaluate the
DC current itself.
}

The manuscript is organized as follows.
{In section II
we present the different mechanism capable of yielding a Rabi coupling,
In section III we present our microscopic modeling 
of the experimental system\cite{Baumann_Paul_science_2015},
and the symmetry difference that allows to distinguish between
the exchange and crystal field mechanisms.
Finally, in
section IV we summarize our conclusions.}

\section{Electrically driven spin excitation mechanisms}

When a voltage difference $V_{\rm RF}(t)$
is applied across the gap   between
the tip and the sample    the electric field
induces a small vertical displacement of
the surface atom $z(t)$ (see Fig. 1b).
We can Taylor expand the spin
Hamiltonian of the surface atom  around $z=0$, the surface atom equilibrium position: 
\begin{equation}
{\cal H} \approx {\cal H} _0  + z(t)  \frac{\partial {\cal H}}{\partial z} \Bigr|_{z=0} 
\label{exp}
\end{equation}
{The previous Hamiltonian consists on a time independent
term $\mathcal{H}_0$ and a time dependent term $z(t)\frac{\partial {\cal H}}{\partial z}
\Bigr|_{z=0}$. The first term determines the excitation spectra of the
quantum system.
When the second term is modulated at a frequency that matches
the energy splitting $\Delta$
between a given pair of eigenstates 
($|M\rangle$ and $|N\rangle$)
of the atomic spin
Hamiltonian  $\mathcal{H}_0$,
transitions}
will be induced provided that
the Rabi force
$\mathcal{F_{N,M}} = \langle 
N|\frac{\partial {\cal H}}{\partial z}|M\rangle \neq 0$. 
Several terms in the Hamiltonian can yield a non-zero contribution
to the term 
$\frac{\partial {\cal H}}{\partial z}$, whose physical meaning
is related on how the Hamiltonian felt by the surface spin changes under small
displacements of the surface atom, and is the responsible of coupling
the surface spin to an electrical signal.

\subsection{Exchange driven mechanism}

Here we propose that
the variation on the tip-surface distance provides 
such coupling in the form of 
exchange interaction
$\mathcal{H}_J = J
(z(t)) \vec{S}_T\cdot\vec{S}$.
Such contribution, already  studied theoretically and observed experimentally
\cite{Schmidt_Lazo_nanolet_2009,Tao_Stepanyuk_prl_2009,PhysRevLett.106.257202,Yan_Choi_natnano_2015,muenks2016correlation,berggren2016electron},
 will be present in any surface spin when probed with a spin
polarized STM and therefore represents a universal
mechanism for electron paramagnetic resonance of individual adatoms.

We ignore  the  quantum fluctuations of the magnetic
moment of the apex atom, quenched  by the combination of an applied magnetic field and
strong Korringa damping with the tip electron bath.  Therefore, we treat the tip
spin in a mean field or classical approximation, following Yan et
al.\cite{Yan_Choi_natnano_2015}, and replace $\vec{S}_T$ by its statistical
average $\langle \vec{S}_T\rangle$.  
For the sake of simplicity, we will restrict the discussion
to the case
when the dynamics is restricted
to the two lowest energy states $|0\rangle$ and $| 1\rangle$, although our description of the atomic spin  states
includes hundreds of multi-electron configurations, as we describe below
\cite{ferron15,Ferron_Lado_prb_2015}.
Within the two-level approximation, 
the relevant operator for the Rabi force associated with
the exchange interaction
is
\begin{equation}
{\mathcal{F}}_J = \frac{\partial J(z)}{\partial z} 
\langle\vec{S}_T  \rangle \cdot \langle 0 |\vec{S}|1 \rangle
\label{rabi_ex}
\end{equation}
This mechanism does not rely on the
specifics of the crystal field of the adatom nor on
spin-orbit coupling, suggesting the possibility
to apply the single spin STM-ESR to a variety of systems,
including S=1/2 atoms and light element magnetism.\cite{gonzalez2016atomic}
\begin{figure}[t]
 \centering
  \includegraphics[width=0.95\columnwidth]{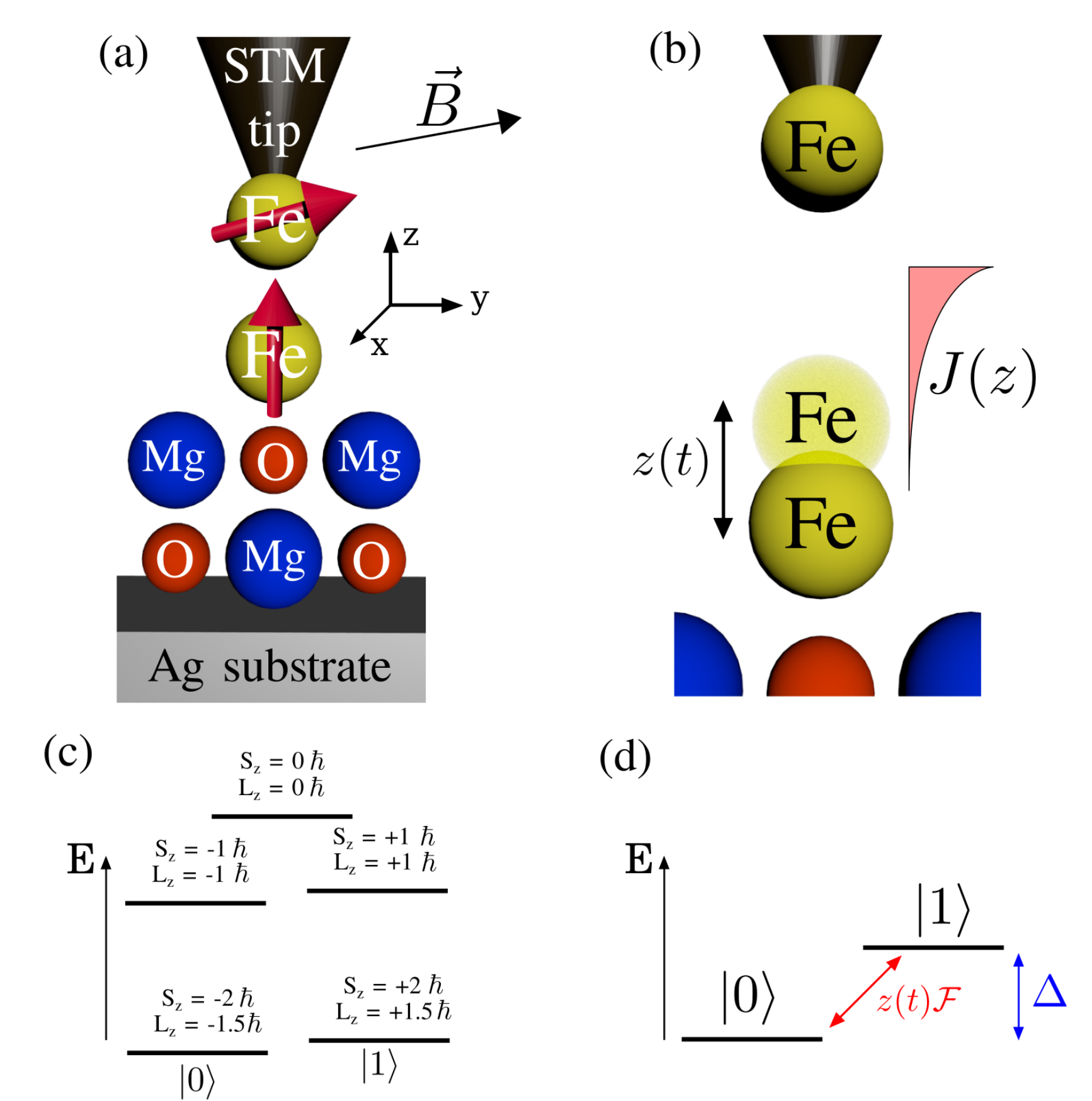}
\caption{ (a) Sketch of an Fe atom in
MgO, where the ESR signal will be measured with
an STM tip. The electric field created by the tip
moves the Fe atom upwards and downwards as shown
in (b), in particular changing the 
the exchange interaction $J(z)$.
Panel (c) shows a sketch of the energy levels of Fe on MgO in the
presence of a small off-plane magnetic field ($B_z $ = 0.2 T).
Panel (d) shows the reduced
low energy Hamiltonian, restricting the dynamics to the two lowest
energy levels in (c).}
\label{fig:sketch}
\end{figure}

\subsection{Crystal field driven mechanism}

A different  mechanism  was proposed by Baumann {\em et al.}  \cite{Baumann_Paul_science_2015}
 specific for  Fe on the (100) MgO surface, where the combination of crystal field and
spin-orbit interaction would couple the two lowest energy
levels of the atomic spin.   The relevant Rabi force for this crystal field mechanism reads
\begin{equation}
\mathcal{F}_{CF} = 
\frac{\partial F_W}{\partial z}
\langle 0 | l_x^4 + l_y^4 | 1 \rangle
\label{rabi_cf}
\end{equation}
where $F_W$ is a crystal field parameter, $l_x$ and $l_y$ are the single-particle orbital momentum operator for the $d$ electrons of  the surface Fe.
%

\subsection{Dipolar coupling driven mechanism}

A third mechanism is provided by the dipolar interaction
between the magnetic moment of the tip and the Fe atom on the surface,
where the tip creates an $z$-dependent in-plane magnetic field,
giving rise a dipolar mixing term of the form
\begin{equation}
\mathcal{F}_{\text{dip}} = 
\mu_B \frac{\partial B_x^{\text{Tip}}}{\partial z} \langle 0 |L_x + 2S_x | 1 \rangle
\label{rabi_dip}
\end{equation}
where $B^{\text{Tip}}_x=2\frac{\mu_0}{4\pi} \frac{\mu_B
S^{Tip}_{x}}{d^3_{\text{Fe-Tip}}}$ is the magnetic field created by the tip,
assuming the moment of the tip lies in-plane
and
$d^3_{\text{Fe-Tip}}$ is the tip-Fe distance.

\subsection{Mechanical coupling between electrical signal and local spin}

{ 
For  the three mechanisms,
the resulting two-level system 
in the $| 0 \rangle, | 1 \rangle$ subspace,
can be written as
$
\mathcal{H}_{\text{eff}} = 
\begin{pmatrix}
\frac{\Delta}{2} & \hbar\Omega(t) \\
\hbar\Omega(t) & -\frac{\Delta}{2} 
\end{pmatrix}
$,
with 
$\Delta=\hbar\omega_0$ the splitting between the ground state
and the first excited, and } the driving Rabi term
\begin{equation}
\Omega(t)= \frac{\mathcal{F}}{\hbar} z(t)
\label{rabi}
\end{equation}
Due to the
oscillating electrical signal applied,
$z(t)=z_0 \cos \omega t$, with $z_0\propto V_{RF}(t)$, we
can write down $\Omega(t)=\Omega_0 \cos \omega t$. We refer to
$\Omega_0=\mathcal{F} z_0/\hbar$ as
the Rabi frequency, which quantifies the efficiency of the driving mechanism
and determines the Rabi time $\tau=\pi/\Omega_0$.   
 This driving force competes with the spin relaxation, characterized by the
energy relaxation and quantum phase 
relaxation times, $T_1$ and $T_2$,  as
described by  the Bloch equations\cite{Abragam_Bleaney_book_1970}. 

Within this approximation, the  steady state solution for the population difference between the ground and excited states is given by the resonance curve 
$
P_0-P_1=\tanh\left( \frac{\hbar\omega_0}{2 k_B T}\right)
\left ( 1 -\frac{\Omega_0^2T_1T_2}{1+ 
 (\omega-\omega_0)^2T_2^2+ \Omega_0^2 T_1 T_2}
\right )
$
where $P_1$ and $P_0$ are the occupation of the ground and excited state,
respectively,
$k_B$ Boltzmann's constant, $T$ the temperature
and the first term on the right-hand side is the thermal equilibrium solution.
Together with $T_1$ and $T_2$,  $\Omega_0$ is critical to assess
by how much $P_1-P_0$ departures from its equilibrium value.
The detection of the previous population imbalance can be
accounted for a magnetoresistive mechanism between the surface
atom and the magnetic tip,although non equilibrium
effects could be relevant and may deserve future attention.\cite{berggren2016electron}

\section{Microscopic modeling}

We now elaborate on the microscopic nature of the operator
$\mathcal{F} = \frac{\partial {\cal
H}}{\partial z}$ relevant for the three mechanisms under discussion. For
 the crystal field mechanism\cite{Baumann_Paul_science_2015} 
, 
 the  vertical displacement of the Fe adatom, $z(t)$   modifies
the crystal field created by the 4 closest Mg ions on the surface
(see Fig. 1b), which in turn modulate the quartic term that allows direct mixing between
$L_z=\pm 2$ \cite{Baumann_Donati_prl_2015} .
The interatomic exchange, that arises from the overlap of the tails of the
atomic orbitals,   decays exponentially  but can be very large at short
distances.
In the following we parametrize 
$J(z)= J_0 e^{-z/{\ell}}$, with $\ell = 0.06$ nm \cite{Yan_Choi_natnano_2015}
and $J_0 = 2$ meV \cite{PhysRevLett.106.257202,Yan_Choi_natnano_2015},
assuming the tip-Fe distance is $d_{\text{Fe-Tip}}=0.6$ nm
and $\langle \vec{S}_T\rangle = 2\hbar$.

The RF bias can induce a Rabi oscillation
by means of the three different mechanisms,  crystal field
(Eq. \ref{rabi_cf}), exchange (Eq. \ref{rabi_ex}) and dipolar
(Eq. \ref{rabi_dip}). The Rabi frequency for all mechanisms depends on
the amplitude of the oscillation (Eq. \ref{rabi}), that is modulated by the
AC bias. Therefore, the mechanism responsible of the oscillation
can be determined by comparing the relative sizes of the
Rabi forces  $\mathcal{F}_J$, $\mathcal{F}_{\text{CF}}$
and $\mathcal{F}_{\text{dip}}$. 
For the exchange mechanism, the exponential
dependence of the exchange coupling implies that
the prefactor in Eq. \ref{rabi_ex} takes a value
$
\frac{\partial J(z)}{\partial z} 
\langle\vec{S}_T \rangle = 66.7 
$
meV/nm. The prefactor for the dipolar
mechanism as given by Eq. \ref{rabi_dip}
yields
$
\mu_B \frac{\partial B_x^{\text{Tip}}}{\partial z} 
= 0.02
$ meV/nm, much smaller than the exchange mechanism and therefore
negligible for typical Tip-Fe distances for ESR. Finally,
the prefactor for the crystal field mechanism
in Eq. \ref{rabi_cf}
requires knowledge of the local crystal field of Fe.
For that matter, we need a spin Hamiltonian for Fe on MgO and, importantly, how
it depends on $z_0$. 

\subsection{First principles calculation of the spin Hamiltonian}

 We derive it starting from a 
density functional theory (DFT) calculation\cite{QE-2009}
for
the system,  following  the same procedure
described in previous works\cite{ferron15,Ferron_Lado_prb_2015}.
 We build a few level model for the electrons in the $d$ orbitals of Fe,
including the crystal field, spin-orbit coupling and electron-electron
interaction and we solve it by numerical diagonalization. The crystal field
part of the Hamiltonian is obtained from the representation of the DFT
Hamiltonian in the basis of maximally localized Wannier
orbitals\cite{RevModPhys.84.1419,mostofi2008wannier90}
$
{\cal H_{\text{CF}}}(z)= D_W(z) l_z^2 + 
F_W(z)
(l_x^4+l_y^4)
$
where $D_W(z) $ and $F_W(z)$ are crystal field parameters that depend on
the vertical coordinate of the surface Fe atom $z$.
 On top of that we add the spin-orbit coupling operator $
\mathcal{H}_{SOC} = \lambda_{\text{SOC}} \vec{l} \cdot \vec{s}$
with $\lambda_{SOC} = 35$ meV,
the Zeeman term
$\mathcal{H}_{B} = \mu_B\vec{B} \cdot (\vec l + 2\vec{s})$
and the electron-electron Coulomb interaction in the $d$ shell. 

\begin{figure}[t!]
 \centering
  \includegraphics[width=0.95\columnwidth]{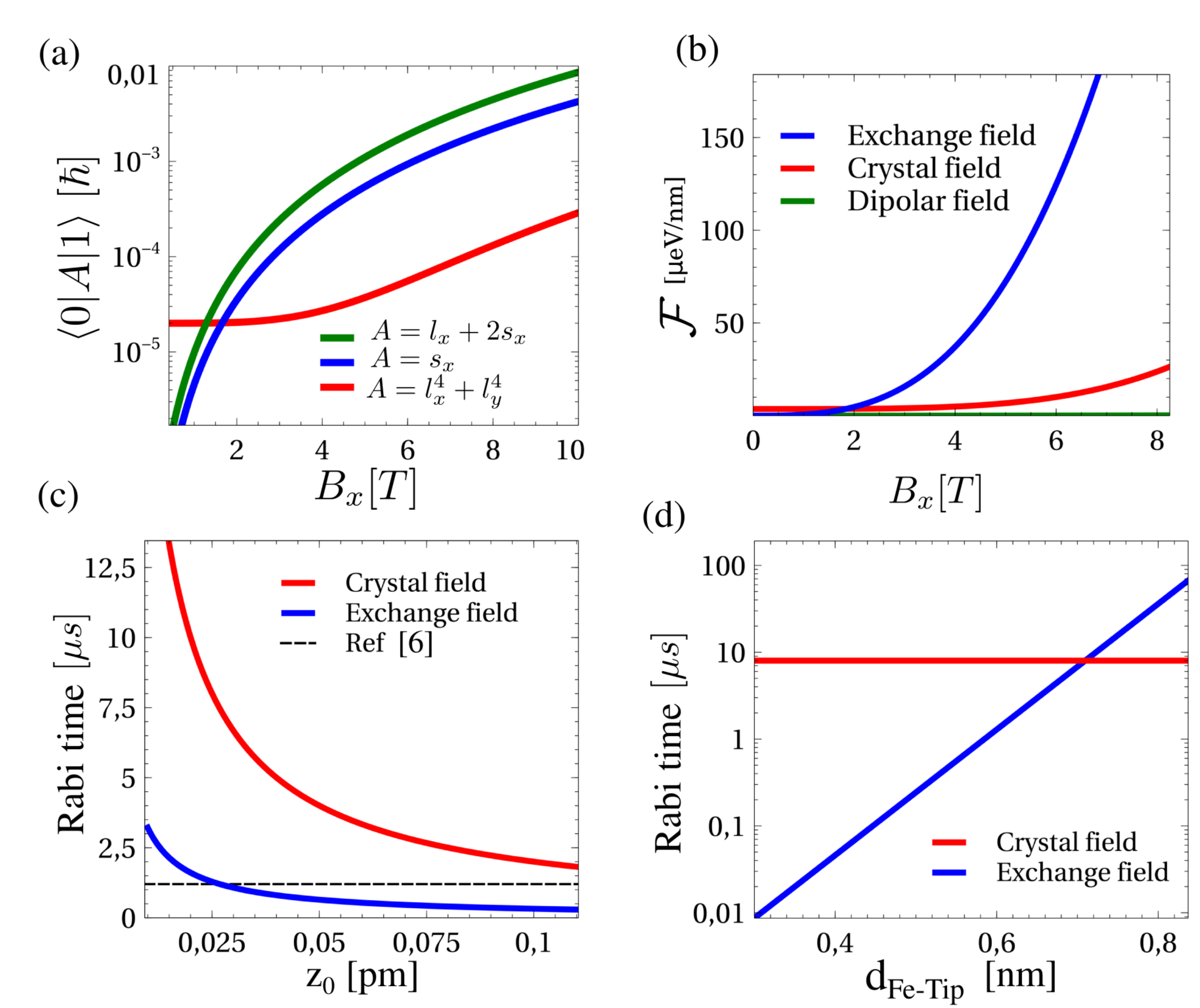}
\caption{ (a) Matrix elements between the ground and first excited state
for the square field perturbation, in-plane Zeeman perturbation
and in-plane exchange perturbation.
With the previous matrix elements and the
dependence of the Hamiltonian with $z$, the Rabi force
can be calculated (b), yielding that the
strongest contribution is the exchange mechanism.
The Rabi time can be calculated from the Rabi frequency Eq. \ref{rabi}
provided the displacement $z_0$ is known, shown in (c). The exponential
dependence of the exchange field produces that the exchange Rabi time
depends on the Fe-Tip distance, while the crystal field mechanism
is assumed independent (d).
}
\label{fig:comparison}
\label{fig:fig2}
\end{figure}

Importantly, the Wannierization
procedure permits to compute both the coefficients
$D_W$ and $F_W$ and 
how they change with the vertical Fe displacement $z$. 
For equilibrium $z=0$  we
find $F_W=-10$ meV \footnote{Sign change in $F_W$ is equivalent to
a 45-degree rotation}
,  $D_W=-290$ meV\footnote{Small variations in $D_W$ do not
influence the low energy spectra},
$\frac{\partial F_W}{\partial z}|_{z=0} = 280 $ meV/nm.
In particular, we also find that
the two lowest energy states $|0\rangle$
and $| 1 \rangle$ of the Hamiltonian eigenstates of
 ${\cal H}_0 \equiv {\cal H}(z)$ have a strong overlap with the states with
quantum numbers  $|L=2,S=2,L_z=\pm 2,S_z=\pm 2\rangle$. This is why the energy
difference $\Delta$ is very sensitive to the application of an off-plane field
$B_z$, and quite insensitive to in-plane components, $B_x,B_y$.

 The results obtained with our method confirm the phenomenological Hamiltonian
describing the low energy  multi-electronic  states for 6 electrons in the $d$
levels of Fe, in the crystal field of the MgO(100) surface, proposed by
  Baumann   {\em et al.} \cite{Baumann_Paul_science_2015,Baumann_Donati_prl_2015}. In
particular, the low energy sector of the Hamiltonian can be parametrized with
${\cal H}(z)= D L_z^2 + F (L_+^4+L_-^4) +  \Lambda\vec{L}\cdot\vec{S} +
 \mu_B\vec{B}\cdot\left(\vec{L}+2\vec{S}\right)
$
where   $L_a$ are the many-body angular momentum operators in the subspace
$L=2$, $\vec{S}$ are spin operators in the $S=2$ subspace,  in both cases
complying with atomic Hund's rules.
By fitting the energies and orbital expectation values
of the lowest 5 states between the multiplet and spin Hamiltonians,
we find the relations $D= -160 $ meV,{\footnote{This effective value 
can show variations if charge fluctuation was considered due to
coupling with the underneath oxygen $p_z$ orbital, giving 
a more accurate comparison with the XMCD spectra in Ref.\cite{Baumann_Donati_prl_2015}}}
$F= -2$ meV and $\Lambda = -11$ meV,
that permit to connect the DFT calculation with the spin model in a simple
manner.

\subsection{Calculation of Rabi matrix elements}

The derivation of the atomic spin Hamiltonian from DFT permits to compute
the relevant matrix elements for the three mechanisms
(Fig. \ref{fig:fig2}a), as well as the Rabi forces (Fig. \ref{fig:fig2}b).
The matrix elements in Fig. \ref{fig:fig2}a
show that the biggest off-diagonal terms correspond
to the spin operator $s_x$ rather than the crystal field operator
$l_x^4 + l_y^4$ at finite in-plane magnetic fields,
giving an advantage to the exchange over the
crystal field mechanism.
When the full Rabi force is calculated,
Fig. \ref{fig:fig2}b, it is obtained
that the exchange remains the leading mechanism,
followed by the crystal field, whereas the dipolar
contribution is nearly negligible.
The role of the 
in-plane magnetic field (applied along the x-axis)
is to mix the wave functions of
$|0\rangle$ and $|1\rangle$  with eigenstates with different $S_z$,  which 
finally enables the transitions between them.


The actual value of the Rabi frequency depends on the magnitude of the
displacement $z_0$.
First, we take the value of $z_0$ as a free parameter and show how the
Rabi time depends on it. In Fig. \ref{fig:fig2}c 
we show the Rabi time as a function
of the Fe displacement $z$ for the crystal field and exchange mechanisms,
as well as the experimental value as a dashed line. The value of the
Fe displacement that would yield a Rabi time comparable to the experimental
one {of $1.2 \mu s$\cite{Baumann_Paul_science_2015}}
would be around $z=0.025$ pm.

\subsection{Mechanical modulation of the Fe position}

In the following we estimate
the magnitude of the vertical displacement assuming
equilibrium between the electric force $ F_{\rm el}= q_{\rm atom} E_{\rm
RF}(t)$ and the spring constant $F_{\rm res}= -kz$, where $k$ is the
restoring force. As the driving frequency is in the GHz range, much smaller
than the standard frequency of  stretching modes,
$\omega_{\text{Fe}}=\sqrt{\frac{k}{M_{\rm
Fe}}}$,  in the THz, we can assume that the atom is always at the
instantaneous equilibrium position
 \begin{equation}
 z(t)= 
\frac{q_{\rm atom} }{k}\frac{ V_{\rm RF}(t)}{d}
 \label{deltaz}
 \end{equation}
with $d$ the decay distance of the electric field, on the order of the
Fe-Tip distance.
DFT calculations yield a value of $k\approx 600$ eV/nm$^2$, that
for $q=2e$
(Fe$^{2+}$), $d=d_{\text{Tip-Fe}}=0.6$ nm and 
the experimental $V_{RF}=8$ meV \cite{Baumann_Paul_science_2015}
gives a value for the Fe
displacement of $z=0.044$ pm, comparable with the one needed for
the Rabi time associated with the exchange field 
\footnote{It must be noted that, although the zero point spread $\left(\frac{\hbar}{M_{\rm Fe}\omega}\right)^{1/2}\simeq 8{\rm pm}$,  is larger than the 
piezoelectric displacement, it does not efficiently couple to the spin because 
the oscillation frequency of 
$\langle z(t) z(0)\rangle$, given by  $\hbar \omega_{Fe}$, is 3 orders of magnitude larger than the spin resonance frequency}
.
For reference, the iron oxide Young modulus
\cite{nicholls1994hardness}
would give $z=0.11$ pm.
We note that the previous estimate would show sizable variations
if non-integer charging of Fe ($q\le 2$) or a larger
voltage drop length ($d>d_{\text{Tip-Fe}}$ nm) are considered. 
%

\begin{figure}[t]
 \centering
  \includegraphics[width=0.95\columnwidth]{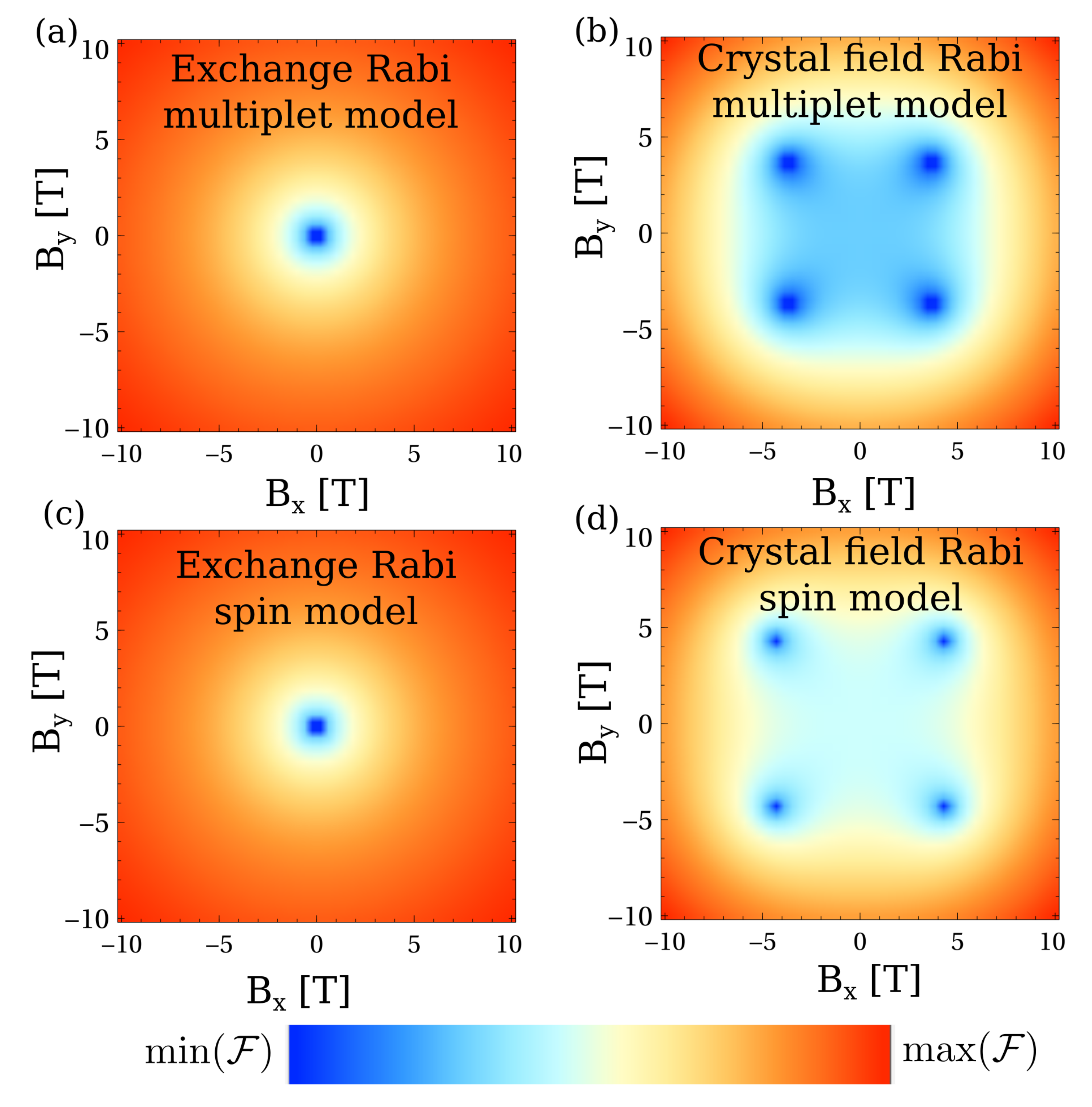}
\caption{Rabi forces
for the exchange
(a,c) and crystal field (b,d) ESR mechanisms,
as a function
of the two components of the in-plane field.
Panels (a,b) correspond to the
multiplet model
and panels (c,d) to the spin model. 
The crystal field contribution shows nodes
when the magnetic field is applied in the (11) directions,
whereas the exchange mechanism is isotropic.
Such symmetry difference allows to determine
the leading mechanism experimentally.
}
\label{fig:fig3}
\end{figure}

\

\

\subsection{Symmetry properties of the Rabi frequency}

We now propose an experimental test to infer which of the proposed  mechanisms is actually driving the spins
in the case of Fe on MgO, based on the dependence of the Rabi energy on the orientation of the in-plane magnetic field. 
For that matter, we plot the Rabi forces,  both for the exchange and crystal
field mechanisms, as a function  of $B_x$ and $B_y$
(Fig. \ref{fig:fig3}). For reference, the states are computed both with the
spin model and with the full multiplet calculation. 
For the exchange
mechanism we obtain a quite isotropic behavior,  expected from the scalar nature of this interaction. 
In contrast,  for the crystal field mechanism, 
the map  reflects the $C_4$ local symmetry,
developing nodes in the (11) direction (Fig. \ref{fig:fig3}b,d).
This same analysis could be carried out to  distinguish 
the relevant mechanism for other atoms deposited on different surfaces.
In case the
dominant mechanism is crystal field,
this makes possible to
determine the local distortions by observing the symmetry
of the Rabi response with the in-plane magnetic fields. 

\section{Conclusion}

We have shown that a modulated exchange coupling between a surface atom
and a magnetic STM tip is an efficient mechanism to induce
electron paramagnetic resonance in the surface spin, by inducing
a Rabi oscillation between the two lowest states. 
Based on DFT and multiplet
calculations, we  show that this mechanism is necessary to 
account for the Rabi time measured in the Fe on MgO experiment\cite{Baumann_Paul_science_2015}. 
Importantly, the exchange driven
mechanism 
shows that the ESR technique is way
more general and could be realized in systems that do not have
a specific crystal field and a sizable spin-orbit coupling.

\section{Acknowledgments}
JLL and JFR  acknowledge financial support by Marie-Curie-ITN
607904 SPINOGRAPH and  COMPETE 2020- FCT  Portugal PTDC/FIS-N
AN/4662/2014Ó (016656). 
JFR acknowledges financial support  by MEC-Spain
(FIS2013-47328-C2-2-P).
AF acknowledges CONICET (PIP11220150100327) and FONCyT (PICT-2012-2866).
We thank F. Delgado for useful discussions. We specially thank
A. Heinrich, C. Lutz and K. Yang for very useful feedback on their
experiment.

\section*{Appendix}

\subsection{Dependence of the resonance frequency with the
STM tip distance}

Our theory can also account for the experimental observation of small
variations of the resonance frequency
   on the tip-Fe $d$ distance for Fe on MgO\cite{Baumann_Paul_science_2015} that have a non-monotonic dependence. 
   Whereas the driving force are dominated by the in-plane component of the tip
magnetic moment, 
   the  much smaller off-plane component adds to the  effective magnetic field
${\cal B}_z(d)$ , that controls 
    the resonant frequency of the surface spin, on  account of its  large
off-plane magnetic anisotropy\cite{Baumann_Paul_science_2015}. 
    This effective field $B_z^{\text{eff}}$ 
is the sum of the actual magnetic field $B_z$ plus  the
exchange and dipolar contributions.

\begin{equation}
B_z^{\text{eff}} = B_z -\frac{\chi J( d_{\text{Fe-tip}}   )}{\mu_B}\langle S^{\text{tip}}_z \rangle + 
\frac{\mu_0}{4\pi}\frac{2\langle \bar S^{\text{tip}}_z \rangle}{d^3_{\text{Fe-tip}}} 
\end{equation}

where $\chi\approx 0.4$ accounts for the fact that exchange coupling
couples only to $S_z$ whereas an effective magnetic field couples
to $2S_z + L_z$. It is worth to note that in the case
of multiple Fe atoms in the STM tip, the effective expectation values
on the dipolar
$\langle \bar S^{\text{tip}}_z \rangle$
 and exchange
$\langle S^{\text{tip}}_z \rangle$
 contribution may be different, since
only the closest atom in the tip
would contribute to exchange interaction, but all of them
to dipolar interaction.

 Whereas tip-surface exchange is
antiferromagnetic\cite{PhysRevLett.106.257202,Yan_Choi_natnano_2015},  dipolar
interaction is ferromagnetic along the off-plane direction.   At short
tip-surface distance, exchange dominates, whereas at longer distance, dipolar
coupling prevails. This competition leads to  of ${\cal B}_{z}$ as a function
of
  $d_{\rm Fe-tip}$,  as seen in the experiment.
%

\subsection{Role of nuclear spin moment in the ESR peaks}

We comment on the role of nuclear spins.   When coupled to a nuclear spin $I$,  the otherwise unique electronic spin resonance splits into $2I+1$ lines, but the total special weight remains the same, so that the visibility of lines is diminished accordingly.   In the
case of Fe, the most abundant isotope (91$\%$)  is $^{56}$Fe, with  $I=0$, which accounts
for the observation of a  single peak\footnote{  We expect that $^{57}$Fe, 2 $\%$ abundance and $I=1/2$, results in 2 peaks}.   In comparison, the only stable  isotope for Co
has $I=7/2$, that will results in 8  peaks split by  the hyperfine coupling, definitely larger than 
the reported line width, but  each of them diminished  by a factor of 8 that  hinders their detection

\subsection{Other mechanisms}
{
We briefly discuss other mechanisms that couple the electric field to the
surface spins.  First, the AC electric field generates an AC magnetic field
that couples to the surface spins.  We have estimated the magnitude of this
field to be in the range of nano Tesla, so that the resulting Rabi time would be
10$^{-6}$ times larger than the one observed experimentally.   
}

{
In reference \onlinecite{berggren2016electron}
a mechanism is proposed based on the
renormalization  of magnetic anisotropy due to exchange interaction with the
tunneling electrons. This renormalization  is a variant of the one observed
experimentally\cite{oberg2014control} for Co atoms on a Cu$_2$N surface as well as
hydrogenated Co atoms on boron nitride  on Rh(111),
\cite{jacobson2015quantum} 
caused by exchange coupling with the surface
electrons.\cite{oberg2014control,delgado2014consequences,jacob2016competition}
The correlation mechanism
proposed in Ref. \onlinecite{berggren2016electron} 
relies on the time modulation of the anisotropy
renormalization induced by the exchange with the tunneling electrons. 
}

{
Finally, in addition to the piezoelectric displacement considered in this work,
the electric field also distorts the electronic orbitals of the surface atom
and tip atom, that should also result in a modulation of their exchange
interaction.  Future work should address the magnitude of this effect. 
}

\bibliographystyle{apsrev4-1}
\bibliography{biblio}{}

\end{document}